\documentclass[twocolumn,british,aps,superscriptaddress,prc,nofootinbib,showpacs,showkeys]{revtex4-1}
\usepackage[T1]{fontenc}
\usepackage[latin9]{inputenc}
\usepackage{color}
\usepackage{babel}
\usepackage{calc}
\usepackage{amsmath}
\usepackage{graphicx}
\usepackage{esint}
\usepackage[unicode=true,pdfusetitle,
 bookmarks=true,bookmarksnumbered=false,bookmarksopen=false,
 breaklinks=false,pdfborder={0 0 1},backref=false,colorlinks=true]
 {hyperref}
\hypersetup{
 pdfborderstyle=,linkcolor=blue,citecolor=cyan}

\makeatletter

\newcommand*\LyXThinSpace{\,\hspace{0pt}}

\usepackage{xcolor}

\makeatother

\begin{document}
\title{Dynamical Electromagnetic fields and Dynamical Electromagnetic Anomaly
in heavy ion collisions at intermediate energies}
\author{Irfan Siddique}
\email{irfansiddique@ucas.ac.cn}
\affiliation{School of Nuclear Science and Technology, University of Chinese Academy
of Sciences, Beijing 101408, China.}

\author{Anping Huang}
\email{huanganping@ucas.ac.cn}
\affiliation{School of Material Science and Physics, China University of Mining
and Technology, Xuzhou 221116, China.}

\author{Mei Huang}
\email{huangmei@ucas.ac.cn}
\affiliation{School of Nuclear Science and Technology, University of Chinese Academy of Sciences, Beijing 101408, China.}


\author{Muhammad Abdul Wasaye}
\affiliation{School of Computer Science and Engineering, Hunan Institute of Technology,
Hengyang 421001, Hunan, China.}

\begin{abstract}
Electromagnetic field produced in non-central heavy ion collisions
play a crucial role in phenomena such as chiral anomalous effects,
directed flow of mesons and splitting of spin polarization of $\Lambda/\bar{\Lambda}$.
A precise description of these fields is essential for quantitatively
studying these effects. We investigate the space-time evolution of the electromagnetic fields by numerically solving Maxwell's equations using the results from the UrQMD model, rather than relying on an ansatz. We present the space-averaged dynamic electromagnetic fields, weighted by energy density, in the central region of heavy-ion collisions. These measurements can serve as a barometer for assessing the effects induced by magnetic fields. Comparing the fields at geometric center of
the collisions, the space-averaged dynamical fields weighted by the
energy density are smaller at the early stage but damp much slower
at the later stage. We discuss the impact of these space averaged
dynamical magnetic fields on the spin polarization and spin alignment
in heavy ion collisions. Additionally, we explore the opportunity
to study non-perturbative regime of Quantum Electrodynamics (QED) by
presenting the simulation results for space averaged dynamical electric
field at intermediate collision energies. Finally, the space-averaged
dynamical electromagnetic anomaly $\text{\textbf{E}}\cdot\boldsymbol{\textbf{B}}$
weighted by energy density is also calculated and compared with experimentally
measured slope parameter $r$.
\end{abstract}
\maketitle

\section{Introduction}

High energy heavy ion collisions provide a unique opportunity to study
matter under extreme temperatures and densities, as well as under
the influence of exceptionally strong electromagnetic (EM) fields.
Simple estimates of the magnetic field strength in non-central heavy
ion collisions, where ions move relativistically, show that the magnitude
can easily reach the hadronic scale $eB\sim m_{\pi}^{2}\sim10^{18}$
Gauss at the Relativistic Heavy ion Collider (RHIC) and even stronger
magnitudes at Large Hadron Collider (LHC) energies \citep{Kharzeev:2007jp,Skokov2009,Asakawa2010,Bzdak:2011yy,Deng:2012pc,Bloczynski:2012en,Tuchin:2013apa,Tuchin:2014iua}.
In high-energy heavy-ion collisions, the extreme temperatures and/or densities result in nuclear matter undergoing a deconfinement phase transition, giving rise to a novel state of matter known as quark-gluon plasma (QGP). During the initial stages of these collisions, a strong magnetic field is generated, which can significantly contribute to the initial energy density and plays a crucial role in the plasma's evolution \citep{Roy2015}. This scenario provides a unique opportunity to explore the interplay between electromagnetic fields and strongly interacting quark and nuclear matter.
The influence of strong magnetic fields on QGP has led to the discovery of several novel phenomena,
such as chiral magnetic effects, charge dependent directed flow, directed
flow of $D^{0}$ mesons, spin polarization of hyperons, splitting
of spin polarization of the $\Lambda/\bar{\Lambda}$ etc \citep{STAR:2021mii,STARCollaboration2022,STARCollaboration2017,Kharzeev:2015znc,Li2020,Huang2016,Zhang:2022lje}. The electromagnetic field also have interesting effects in pair production \citep{Sheng:2017lfu,Sheng:2018jwf} and transport properties \citep{Fukushima:2015wck, Hattori:2017qih, Lin:2021sjw, Peng:2023rjj}. 

The effects of magnetic fields and related observables in heavy ion
collision depends on the strength and evolution of these fields. Particularly to induce an effect on observables, the magnetic fields needs to be strong enough and sufficiently long lived in the evolving matter in heavy ion collisions. While the initial spatial distribution and strength can be accurately calculated at the beginning of heavy ion
collisions, the situation becomes increasingly complex over time as
the system evolves through various phase transitions. This complexity
makes precise calculations challenging at later stage. Typically,
the simulation of electromagnetic fields begins by determining the
charge density of the colliding nuclei, which is often achieved using
the Woods-Saxon distribution or by sampling charge positions within
the nucleus via the Monte Carlo Glauber model 
\citep{Bzdak:2011yy,Bloczynski:2012en,Siddique2022}.
The nucleons are then boosted in opposite direction (usually $\pm z$
directions) and allowed to collide. The simplest approach assumes
that the two colliding nuclei pass through each other without further
interaction. However, more sophisticated and realistic methods involve
using transport models such as UrQMD (Ultra Relativistic Quantum Molecular
Dynamics) \citep{Bass:1998ca,Bleicher:1999xi}, AMPT (A multiphase
transport model) \citep{Lin:2004en,Lin2021} etc. These models help
in simulating the entire collisions process and provide a more detailed
framework for calculating the evolution of electromagnetic fields. 

To calculate the space-time evolution of electromagnetic fields, one
generally assumes either the absence of a medium (vacuum scenario)
or the presence of the medium such as QGP. Previous studies have shown
that in the vacuum scenario, magnetic fields decays rapidly with time,
following $t^{-3}$ dependence during the early phases of evolving
matter \citep{Huang2016,Hattori2017}. However, the time evolution
of these fields can be significantly modified when considering the feedback from QGP medium. Specifically, the decay of the magnetic field is significantly slowed when the feedback from the quark-gluon plasma (QGP) is taken into consideration.
\citep{McLerran:2013hla,Tuchin:2013apa,Tuchin:2014iua,Li:2016tel,Siddique:2019gqh,Siddique:2021smf,Siddique2022,Anping2023}.
To describe the induced effects of electromagnetic fields, it is crucial
to understand the quantitative behavior of electromagnetic fields
using a more realistic approach. This requires a framework that incorporate
the entire collision process and solves the Maxwell's equations in
the presence of a conducting medium such as QGP, where induced Faraday
currents significantly slows down the decay of produced fields. In
this work instead of relying on ansatz, we simulate the full collision
process using the UrQMD model to calculate the corresponding currents.
We then numerically solve Maxwell's equations, incorporating finite
conductivities, to obtain the dynamical electric and magnetic fields.
We also provide a comparison with the vacuum scenario. Furthermore,
we also extend our calculations to evaluate electromagnetic anomaly
$\text{\textbf{E}}\cdot\boldsymbol{\textbf{B}}$ from the dynamical
electric and magnetic fields. 

We present numerical results for collision energy in the range of $3.5\text{ GeV}\leq\sqrt{s_{NN}}\leq27\text{ GeV}$ in this work. It is noticed that the dynamics of the system differ significantly between lower and
higher collision energies,  At lower collision energies, the Landau
picture \citep{Fermi1950,Landau:1953gs} is realised due to the baryon
stopping, where after the collision the colliding nuclei are slowed
down for a moment and ions may stick together and act as whole gigantic
ion which can create strong Columb electric field, which may as well
subsequently influence the evolution of electromagnetic fields produced
in heavy ion collisions. In contrast, at higher collision energies, 
Bjorken picture \citep{Bjorken:1982qr} is realised, where due to
highly relativistic movement, the colliding nucleons penetrate through
each other without significant interaction or sticking together. Moreover,
the lifetime of electromagnetic fields is expected to be longer at
relatively lower collision energies compared to higher collision energies.
The extended lifetime of electromagnetic fields can also play important
role in field induced phenomenon. 

After providing the brief introduction, in Section \ref{sec:2} we
give the expressions for dynamical electric and magnetic fields in
a system having finite conductivities. In Section \ref{sec:3}, we
provide simulation results and discussions for the dynamical electromagnetic
fields. Finally we summarize our findings in Section \ref{sec:summary}.

\section{\label{sec:2} Calculation of Dynamical electromagnetic field}

\subsubsection*{A. Solving Maxwell's Equations: Numerical Methods and Approaches}
The covariant form of Maxwell's equation with external sources, given
by the current density four-vector $j^{\mu}=\text{\ensuremath{\left(\rho,\boldsymbol{j}\right)}}$
is: 
\begin{equation}
\begin{aligned}\partial_{\mu}F^{\mu\nu} & =j^{\nu},\\
\partial_{\mu}\tilde{F}^{\mu\nu} & =0,
\end{aligned}
\end{equation}
where the field strength tensor of electromagnetic field is $F^{\mu\nu}=\partial^{\mu}A^{\nu}-\partial^{\nu}A^{\mu}$
and it's dual tensor is $\tilde{F}^{\mu\nu}=\frac{1}{2}\epsilon^{\mu\nu\rho\sigma}F_{\rho\sigma}$.
The covariant form of Maxwell's equations can be written in a three-vector
form as 
\begin{equation}
\begin{aligned}\boldsymbol{\nabla}\cdot\boldsymbol{\textbf{B}}= & 0,\\
\boldsymbol{\nabla}\cdot\boldsymbol{\textbf{E}}= & \rho,\\
\boldsymbol{\nabla}\times\boldsymbol{\textbf{B}}= & \partial_{t}\boldsymbol{\textbf{E}}+\text{\textbf{J}}_{3}\left(\sigma,\sigma_{\chi}\right),\\
\boldsymbol{\nabla}\times\boldsymbol{\textbf{E}}= & -\partial_{t}\boldsymbol{\textbf{B}},
\end{aligned}
\label{eq:eq2}
\end{equation}
where $\text{\textbf{J}}_{3}\left(\sigma,\sigma_{\chi}\right)=\text{\textbf{J}}_{ext}+\sigma\boldsymbol{\textbf{E}}+\sigma_{\chi}\boldsymbol{\textbf{B}}$
with $\text{\textbf{J}}_{ext}$ being external current density and
$\rho$ being external charge density. In above equation $\sigma$
is electric conductivity and $\sigma_{\chi}$ is chiral magnetic conductivity
of charged conducting medium (QGP). The first two lines of equation
\ref{eq:eq2} are constraints, while last two lines of equation \ref{eq:eq2}
are used to derive the dynamical electric and magnetic fields. 

To evaluate the dynamical electromagnetic fields numerically, we first
calculate the $j^{\mu}$ using the UrQMD. The UrQMD model simulate
a full collision process, providing the phase-space distribution of
all hadrons in a heavy ion collision event. The positions $\left(x_{n}\left(t\right)\right)$ and
momenta $\left(p_{n}^{\mu}\left(t\right)\right)$ of charged particle
are function of time and are provided by the simulation of UrQMD model.
The electric current at a position $\left(t,x\right)$ can be given
as 
\begin{equation}
\begin{aligned}j^{\mu}\left(t,x\right)= & \sum_{n}\frac{p_{n}^{\mu}\left(t\right)}{p_{n}^{0}\left(t\right)}\rho_{n}\left(t,x\right),\end{aligned}
\end{equation}
where $n$ labels $n-$th hadron. The charged density of $n-$th hadron
$\left(\rho_{n}\right)$ is localized at $\boldsymbol{x}_{n}$, and
to account for the finite size of the charge distribution, we smear
the charge density by using Gaussian distribution
\begin{equation}
\begin{aligned}\rho_{n}\left(t,x\right)= & \frac{q_{n}\gamma_{n}\left(t\right)}{\left(\sqrt{2\pi}\varrho\right)^{3}}exp\left[-\frac{\left|\boldsymbol{x}-\boldsymbol{x}_{n}\left(t\right)\right|^{2}}{2\varrho^{2}}\right.\\
 & \left.-\frac{\gamma_{n}^{2}\left(t\right)\left(\boldsymbol{v}_{n}\left(t\right)\cdot\text{\ensuremath{\left[\boldsymbol{x}-\boldsymbol{x}_{n}\left(t\right)\right]}}\right)^{2}}{2\varrho^{2}}\right],
\end{aligned}
\end{equation}
where $q_{n}$ is the electric charge, $\varrho$ is the smearing
width and $\gamma$ is the Lorentz factor for the $n-$th hadron.
Once the current density $j^{\mu}$ is calculated with UrQMD we can
numerically solve Maxwell equations by constructing corresponding
wave equations for the electric and magnetic fields.

In order to numerically solve Maxwell's equations we use strategy
as described in \citep{McLerran:2013hla,Anping2023}, we decompose
the electric and magnetic field as 
\begin{equation}
\begin{aligned}\boldsymbol{\textbf{F}}= & \text{\textbf{F}}_{ext}+\text{\textbf{F}}_{int}\end{aligned}
,\label{eq:Fext-int}
\end{equation}
where $\boldsymbol{\textbf{F}}$ denotes either $\text{\textbf{B}}$
or $\text{\textbf{E}}$. In above equation the subscript `ext' denotes
external part of field which is originated by the source contribution
from the fast moving charge particles in heavy ion collisions. The
subscripts `int' represents the induced part of field which is generated
in the QGP. So Maxwell's equations given in Eq. (\ref{eq:eq2}) can
also be separated into external and internal part. The external part
is given as 
\begin{equation}
\begin{aligned}\boldsymbol{\nabla}\cdot\boldsymbol{\textbf{B}}_{ext}= & 0,\\
\boldsymbol{\nabla}\cdot\boldsymbol{\textbf{E}}_{ext}= & \rho,\\
\boldsymbol{\nabla}\times\boldsymbol{\textbf{B}}_{ext}= & \partial_{t}\boldsymbol{\textbf{E}}_{ext}+\text{\textbf{J}}_{ext},\\
\boldsymbol{\nabla}\times\boldsymbol{\textbf{E}}_{ext}= & -\partial_{t}\boldsymbol{\textbf{B}}_{ext},
\end{aligned}
\label{eq:external-maxwell}
\end{equation}
whereas the internal part can be given as
\begin{equation}
\begin{aligned}\boldsymbol{\nabla}\cdot\boldsymbol{\textbf{B}}_{int}= & 0,\\
\boldsymbol{\nabla}\cdot\boldsymbol{\textbf{E}}_{int}= & 0,\\
\boldsymbol{\nabla}\times\boldsymbol{\textbf{B}}_{int}= & \partial_{t}\boldsymbol{\textbf{E}}_{int}+\sigma\left(\boldsymbol{\textbf{E}}_{ext}+\boldsymbol{\textbf{E}}_{int}\right)-\sigma_{\chi}\left(\boldsymbol{\textbf{B}}_{ext}+\boldsymbol{\textbf{B}}_{int}\right),\\
\boldsymbol{\nabla}\times\boldsymbol{\textbf{E}}_{int}= & -\partial_{t}\boldsymbol{\textbf{B}}_{int}.
\end{aligned}
\label{eq:internal-maxwell}
\end{equation}

We use Yee's algorithm \citep{Yee1966}, also known as Finite-Difference
Time Domain (FDTD) method to numerically solve above set of equations
in space and time, details also in \citep{Anping2023}. In FDTD the
components of the electric and magnetic field are staggered in both
time and space, whereas staggering ensures a second order acurracy
on both space and time. For implementation of this algorithm, first
we define 3D space with grid steps $\Delta x,$ $\Delta y,$ $\Delta z$,
and define time step such that it satisfies Courant-Friedrichs-Lewy
(CFL) stability condition i.e., $\Delta t\leq\left(\sqrt{\frac{1}{\Delta x^{2}}+\frac{1}{\Delta y^{2}}+\frac{1}{\Delta z^{2}}}\right)^{-1}$and
then by using current density $j^{\mu}$ in each grid we initialize
the fields. In next step we iterate over time where we update $j^{\mu}$ and
so update fields $\boldsymbol{\textbf{F}}\left(\text{i.e., }\boldsymbol{\textbf{B}}\text{ or }\boldsymbol{\textbf{E}}\right)$
using finite difference form of $\boldsymbol{\nabla}\times\boldsymbol{\textbf{F}}$,
and also ensure that the fields are updated at each time step. In
this way, we can record EM fields at each time step and space grids.
The dynamic electric and magnetic fields are obtained by combining
results from internal and external components by adding them together
as shown in Eq. (\ref{eq:Fext-int}). 

\subsubsection*{B. Space-Average Dynamical Electromagnetic fields weighted Energy Density}

The EM fields produced in heavy-ion collisions are highly inhomogeneous
in space and time \citep{Deng:2012pc,Li:2016tel,Zhao2019a, Zhao:2019ybo,Siddique2022},
evaluating the field at some specific space-time point could lead
to over- or under-estimation of the field induced effects. To mitigate
this, we use space-averaged dynamic EM fields weighted by energy density,
following the fact that the region with lower energy density contribute
less to the chiral effects arising from the electromagnetic field
and matter density (details can also be found in \citep{Siddique:2021smf}).
Since we divide the whole space into small grids so space-averaged
dynamical fields can be given as
\begin{equation}
\begin{aligned}\left\langle \boldsymbol{\textbf{F}}\right\rangle_{E}  & \left(t\right)=\end{aligned}
\frac{\sum_{i}\varepsilon_{i}\left(t\right)\boldsymbol{\textbf{F}}_{i}\left(t\right)}{\sum_{i}\varepsilon_{i}\left(t\right)},\label{eq:space-averaged}
\end{equation}
where $\varepsilon_{i}\left(t\right)$ is the energy density in the
$i-$th grid and $\boldsymbol{\textbf{F}}$ represents the electric
$\boldsymbol{\textbf{E}}$ or magnetic field $\boldsymbol{\textbf{B}}$
in the center of the same grid. We use $\langle \boldsymbol{...}\rangle_{E}$ to represent energy density weighted results in our draft. Here we mention that for energy density
calculations, we only consider the particles in momentum rapidity
range $-0.5\leq\text{Y}\leq0.5$ in the fireball. While calculating
the current densities and solving Maxwell's equation to obtain dynamical
electric and magnetic fields all charged particles from UrQMD are
taken into account. 

\section{Simulation results and Discussions\label{sec:3}}

In this section, we present the simulation results by using the numerical
method discussed in the previous section. We assume that one nucleus
moves along $+z$ direction and other along $-z$ direction, with
their centers located at $x=b/2$ and $x=-b/2$ respectively, where
$b$ is impact parameter and the reaction plane is formed by $xz-$plane.
This setup positions the orbital angular momentum (OAM) direction
along the $-y$ direction. In set of Maxwell's equation, we set the
electric conductivity $\sigma=5.8$ MeV, consistent with the lattice
quantum chromodynamics calculation \citep{Ding2011,Aarts2015}. The
chiral magnetic conductivity is set to $\sigma_{\chi}=1.5$ MeV, corresponding
to $\mu_{5}\sim100$ MeV, as used in earlier studies \citep{Li:2016tel,Siddique2022,Siddique2024}.
Although the electric conductivity varies with time, however the
influence of the time dependent conductivity on the field lifetime during the hydrodynamic evolution
is not very significant \citep{Tuchin2013}. Furthermore, it has been
shown in \citep{Li2023} that the behavior of time-dependent and constant
electric conductivity is similar at both early and late stages of
evolution, and at lower energies this difference becomes minimal.
Thus as a first step, treating conductivities as constant is a good
approximation in using UrQMD model to obtain four-current and solving
Maxwell's equation at intermediate collision energies.

For numerical simulations, we consider Au+Au collisions with a spatial
volume as $-15\text{ fm}\leq x,y\leq15$ fm with $dx=dy=0.5$ fm,
and $-15\text{ fm}\leq z\leq15$ fm with $dz=0.1$ fm. The time step
is set to $dt=0.05$ fm/c. Eqs \ref{eq:external-maxwell}, \ref{eq:internal-maxwell}
and \ref{eq:space-averaged} has been numerically solved by using
aforementioned Yee's algorithm to obtain results in this section. Below,
we present simulation results for the spatial distribution and time
evolution of the dynamical electromagnetic fields.

\subsubsection{Spatial Distributions of electromagnetic fields:\label{subsec:Spatial-Distributions:}}

\begin{figure*}
\noindent\fbox{\begin{minipage}[t]{1\columnwidth - 2\fboxsep - 2\fboxrule}%
\includegraphics[width=8.5cm]{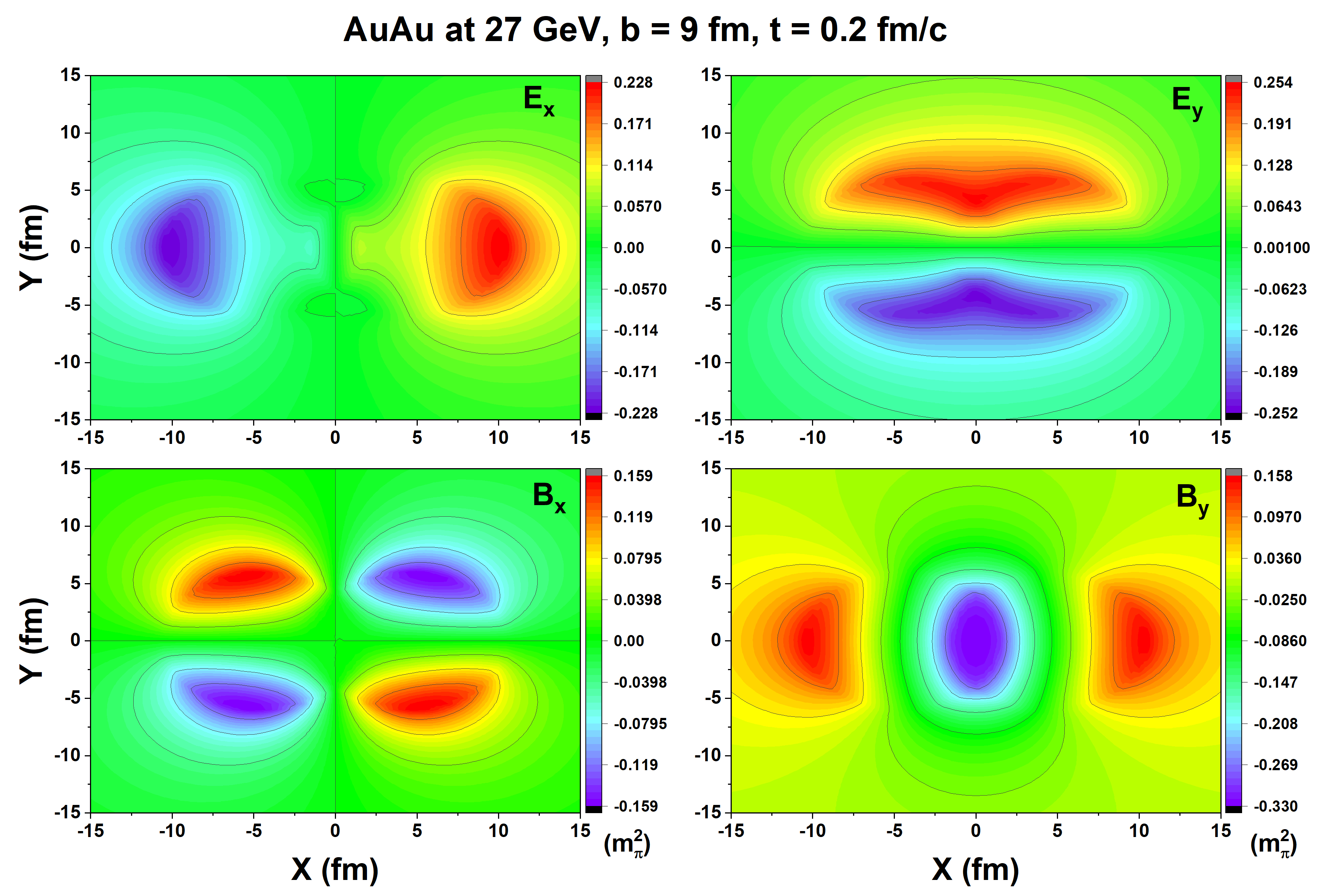}

A\label{A-panel}%
\end{minipage}}%
\noindent\fbox{\begin{minipage}[t]{1\columnwidth - 2\fboxsep - 2\fboxrule}%
\includegraphics[width=8.5cm]{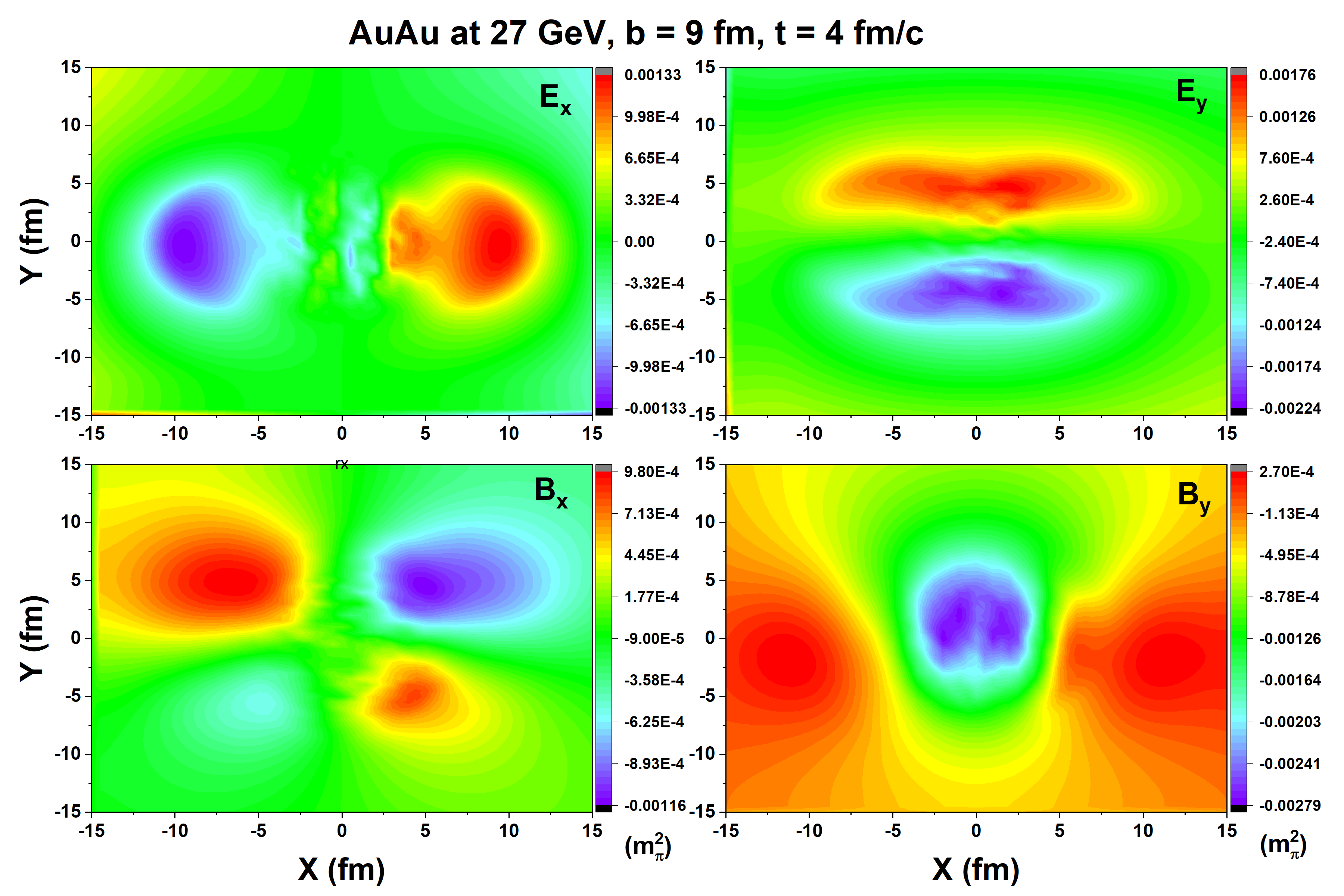}

B%
\end{minipage}}

\caption{\label{fig:contour27-ExyBxy}Spatial distribution of transverse components
of the dynamical electric and magnetic field in transverse plane in
Au+Au collisions at 27 GeV and $b=9$ fm at $t=0.2$ fm/c (panel A) and $t=4$
fm/c (panel B).}
\end{figure*}

In Fig. \ref{fig:contour27-ExyBxy}, we shows contour plots of the
spatial distributions of the dynamical electric and magnetic field
components (in units of $m_{\pi}^{2}$ ) for Au+Au collisions at $\sqrt{s_{NN}}=27$
GeV with $b=9$ fm. These distributions are highly inhomogeneous,
consistent with findings from previous studies \citep{Deng:2012pc,Li:2016tel,Zhao2019a,Zhao:2019ybo,Siddique2022}.
The figure has two panels in Fig. \ref{fig:contour27-ExyBxy}: Panel
A (left) shows the fields at $t=0.2$ \LyXThinSpace{} fm/\ensuremath{c}
, and Panel B (right) shows the fields at $t=4$ fm/c. These two snapshots
illustrate the evolution of the fields over time. In Panel A, the
spatial distribution of the electromagnetic field components is symmetric
because contributions from $\text{\textbf{F}}_{int}$ (the internal
fields) are minimal in $\text{\textbf{F}}_{tot}$. At the early stage,
the dominant contribution comes from the fast-moving source charge
particles, represented by $\text{\textbf{F}}_{ext}$. In Panel B,
the spatial distribution becomes partially asymmetric due to the significant
contribution from $\text{\textbf{F}}_{int}$, which includes the effects
of electric and chiral magnetic conductivities in the QGP. These results
highlight the transition from a purely external field contribution
at early times to a combined contribution from both external and internal
fields as the system evolves. 

\begin{figure*}
\includegraphics[width=10cm]{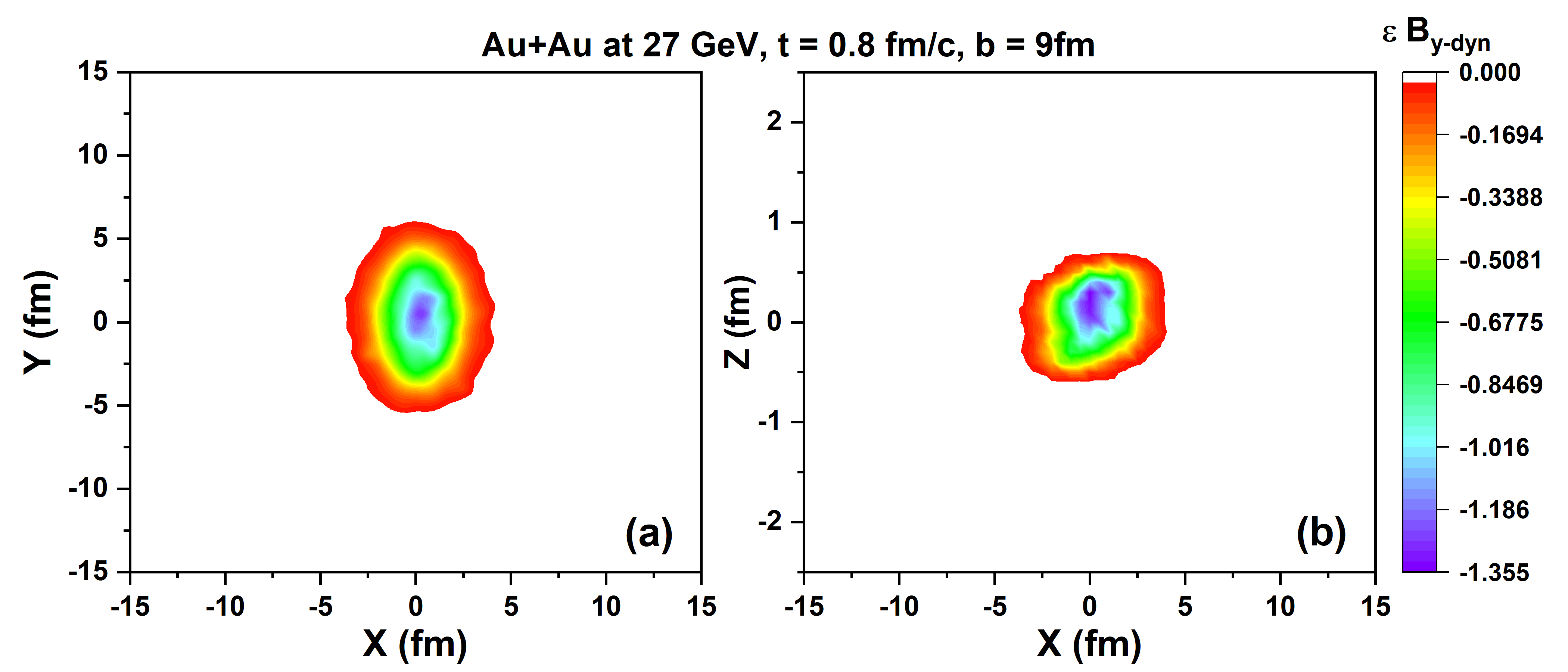}

\caption{\label{fig:spatialEdenBy}Spatial distribution of the product of energy
density and dynamical $eB_{y}$ in transverse (a) and reaction plane
(b) at $t=0.8$ fm/c in Au+Au collisions at 27 GeV and $b=9$ fm.}
\end{figure*}

We further provide spatial distribution of the product of energy density
and magnetic field along the OAM direction $\left(\varepsilon B_{y}\right)$
in Fig. \ref{fig:spatialEdenBy}. This calculation is crucial for
evaluating energy density-weighted fields, which serve as a barometer
for field-induced effects. As described in the previous section, energy
density calculations only include particles within the momentum mid-rapidity
range $-0.5<\text{Y}<0.5$ to minimize the influence of spectators
and boundary regions of quark or nuclear matter. This ensures that
the focus remains on the central region, where $\varepsilon B_{y}$ is
non-vanishing. The spatial distribution of $\varepsilon B_{y}$ clearly
indicates that significant contributions are concentrated in the central
region of the fireball, reflecting the dominance of QGP effects in
this area.

\subsubsection{\label{subsec:Time-evolution:}Time evolution of the magnetic field:}

\begin{figure}
\includegraphics[width=8cm]{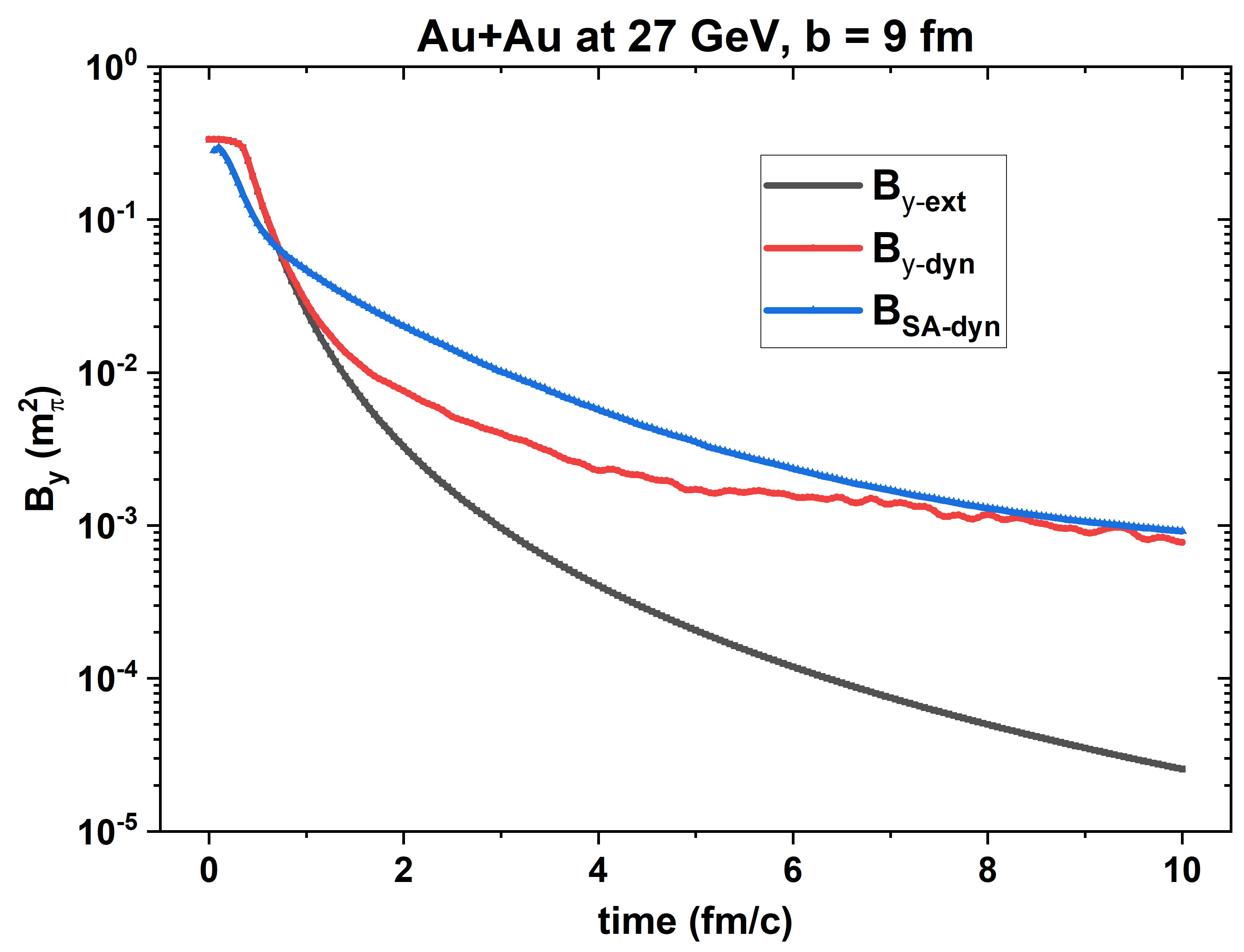}\caption{\label{fig:Au+Au-at-27time}Time evolution of magnetic field \textbf{$B_{y}$}
in Au+Au collisions at $\sqrt{s_{NN}}=27$ GeV and $b=9$ fm. The
comparison has been made for $B_{y-ext}$ at (0,0,0), $B_{y-dyn}$
at (0,0,0) and space averaged dynamical magnetic field weighted by
energy density ($B_{SA-dyn}=\left\langle B_{y-dyn}\right\rangle _{E}$).}
\end{figure}

In this subsection, we discuss the time evolution of the dynamical
magnetic fields. In Fig. \ref{fig:Au+Au-at-27time}, we give the comparison
between vacuum case $\left(B_{y-ext}\text{ at (0,0,0), black line}\right)$,
dynamical magnetic field $\left(B_{y-dyn}\text{ at (0,0,0), red line}\right)$
and space averaged dynamical magnetic field weighted by energy density
$\left(B_{SA-dyn}\text{, blue line}\right)$ in a Au+Au collisions
at $\sqrt{s_{NN}}=27$ GeV and $b=9$ fm. We focus on the time evolution
of only $y-$component of the magnetic field weighted by energy density
$B_{SA-dyn}=\left\langle B_{y-dyn}\right\rangle _{E}$ ($E$ labels
for energy as weight) because all other components of magnetic field
and electric field are almost vanishing because of symmetric collisions
and spatial distributions. 

The dynamical magnetic field $B_{y-dyn}$ (red line) initially behaves
similar to the vacuum scenario $B_{y-ext}$ (black line). However,
at later times, the decay of $B_{y-dyn}$ slowdown due to contributions
from the QGP medium, where the conductivities $\sigma$ and $\sigma_{\chi}$
play a role. This behavior reflects the time required for QGP-induced
effects to build up after the collision. The space-averaged magnetic
field $\left\langle B_{y-dyn}\right\rangle _{E}$ (blue line) exhibits
an interesting trend when compared to the magnetic field at the origin.
At initial times, the magnitude of $\left\langle B_{y-dyn}\right\rangle _{E}$
is smaller. However, at intermediate times, it becomes larger than
fields at origin (red and black line) due to contributions from regions
with higher energy density in the fireball. At later times $t>7$
fm/c $\left\langle B_{y-dyn}\right\rangle _{E}$ and $B_{y-dyn}$
at origin have almost similar magnitude, though $\left\langle B_{y-dyn}\right\rangle _{E}$
remains slightly larger.

\begin{figure*}
\includegraphics[width=14cm]{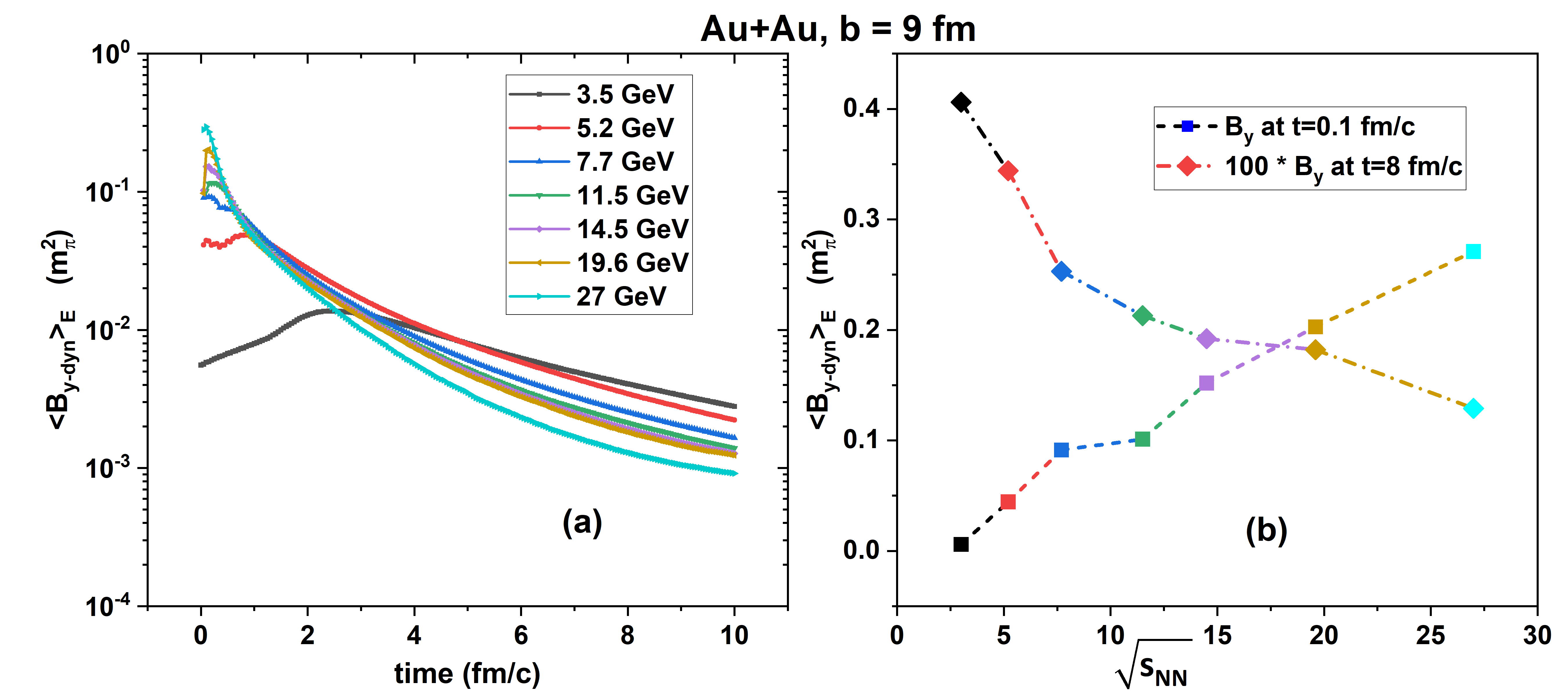}

\caption{\label{fig:Space-averaged-dynamical-magneti}Time evolution of space-averaged
dynamical magnetic field $\left\langle B_{y-dyn}\right\rangle _{E}$
as function of collision energy in Au+Au collisions at $b=9$ fm in
left panel. Right panel shows the magnitude of $\left\langle B_{y-dyn}\right\rangle _{E}$
at early time $\left(t=0.1\text{ fm/c}\right)$ shown by square symbols
and later time $\left(t=8\text{ fm/c}\right)$ shown by diamond symbols. }
\end{figure*}

In Fig. (\ref{fig:Space-averaged-dynamical-magneti}a), we give results
for time evolution of $\left\langle B_{y-dyn}\right\rangle _{E}$
 for different collision energies. The peak value of $\left\langle B_{y-dyn}\right\rangle _{E}$
increases with collision energy, as the production of magnetic fields
is directly related to the movement of charged particles. As collision
energy increases the system have fast movements of charges and higher
energy densities, so the dynamics of the particles such as momentum
and currents they generate also intensify, leading to stronger maximum
magnetic fields. Conversely, the lifetime of the space-averaged dynamical magnetic
field decreases with increasing collision energy. This inverse relationship
arises because as collision energies increases the speed of charges
also increases, which causes quick decay of the magnetic field. 

We also show the magnitudes $\left\langle B_{y-dyn}\right\rangle _{E}$ at
early time $(t=0.1\text{ fm/c, square symbols})$ and late time $(t=8.0\text{ fm/c, diamond symbols})$
in Fig. (\ref{fig:Space-averaged-dynamical-magneti}b). At early times, the magnitude increases with collision energy, while
at late times, the field magnitude decreases as collision energy increases.
This highlights that at lower collision energies, the magnetic field
persists longer and larger magnitude at late time in the mid-central
rapidity fireball system, despite its smaller peak magnitude.

Now we see the impact of the $\left\langle B_{y-dyn}\right\rangle _{E}$
on obseravble quantities, such as the splitting between the global
spin polarization of $\Lambda$ and $\bar{\Lambda}$ following references
\citep{Becattini2017,Mueller2018,Peng2023}. The difference between
the global polarization of $\Lambda$ and $\bar{\Lambda}$ is defined
as \citep{STARCollaboration2017}
\begin{equation}
\Delta\mathcal{P}=\mathcal{P}_{\Lambda}-\mathcal{P}_{\bar{\Lambda}},\label{eq:splitting eq}
\end{equation}
the magnetic field induced global polarization of $\Lambda$ $\left(\bar{\Lambda}\right)$
can be calculated as
\begin{equation}
\mathcal{P}_{\Lambda\text{ \ensuremath{\left(\bar{\Lambda}\right)}}}=\frac{\mu_{\Lambda\text{ \ensuremath{\left(\bar{\Lambda}\right)}}}B_{y}}{T},
\end{equation}
where $\mu_{\Lambda}=-0.613\mu_{N}$ is magnetic moment of of $\Lambda$
hyperon, with $\mu_{N}$ being nuclear magneton and $\mu_{\Lambda}=-\mu_{\bar{\Lambda}}$.
The temperature $T$ marks the point at which the hyperon spin ceases
to evolve and we use $T=155$ MeV. So Eq. (\ref{eq:splitting eq}) can
be given as \citep{Li2023}
\begin{equation}
\Delta\mathcal{P}=0.0826\text{ }\frac{B_{y}}{m_{\pi}^{2}}.
\end{equation}

The numerical results presented in Fig. \ref{fig:Space-averaged-dynamical-magneti}
show that the $\left\langle B_{y-dyn}\right\rangle _{E}$ at initial
and intermediate time has magnitude of order $10^{-2}-10^{-1}$ $m_{\pi}^{2}$,
and the magnitude at late time is of order $10^{-3}-10^{-2}$ $m_{\pi}^{2}$.
The effect of $\left\langle B_{y-dyn}\right\rangle _{E}$ is negligible
on the splitting between the global polarization of $\Lambda$ and
$\bar{\Lambda}$. Our result is consistent with the results reported
in \citep{Li2023,Peng2023} and with recent STAR data \citep{STAR2023}.

\subsubsection{Time evolution of the electric field}

\begin{figure}
\includegraphics[width=8cm]{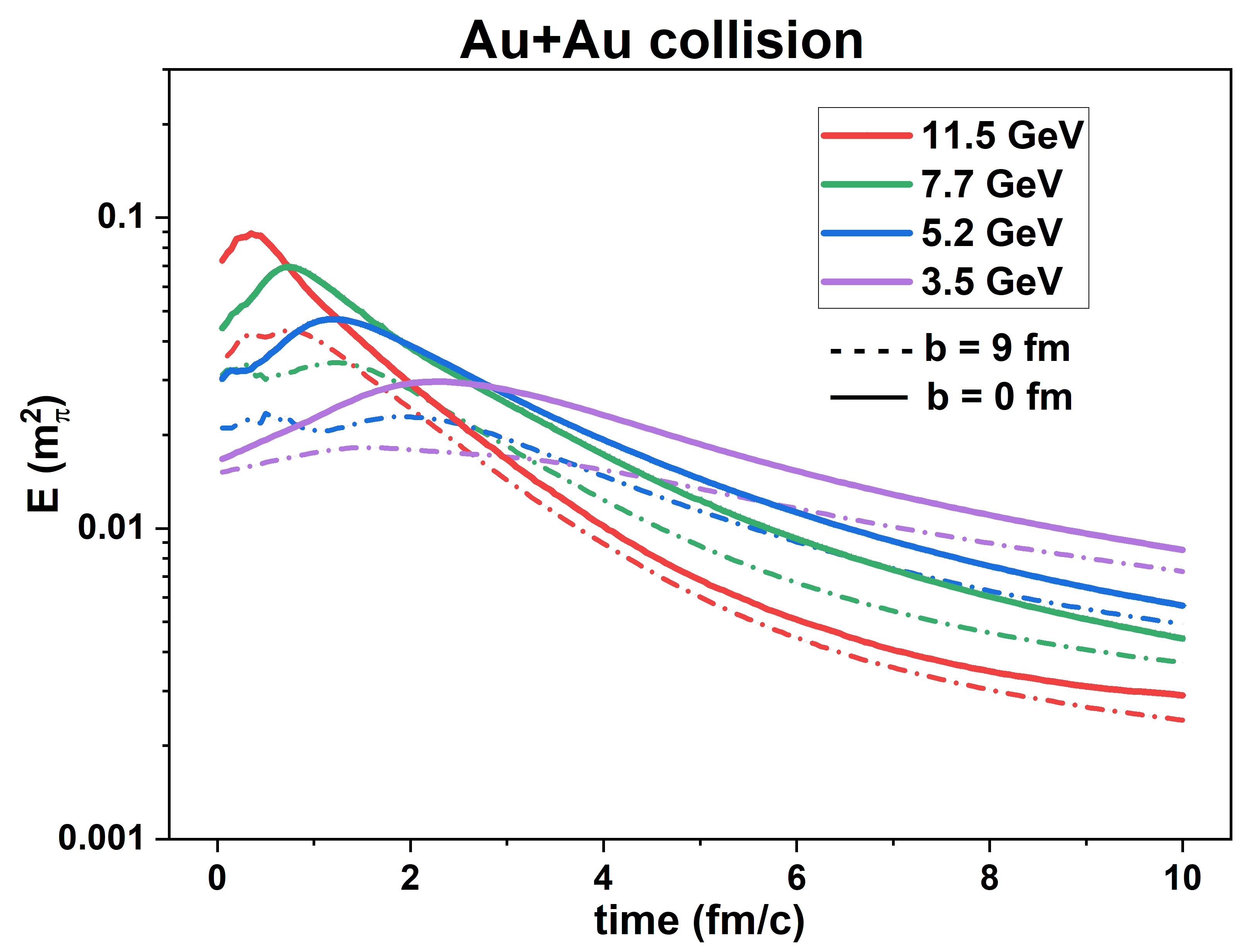}

\caption{\label{fig:Space-averaged-dynamical-electri}Space-averaged dynamical
electric field $\left\langle \boldsymbol{\text{E}}_{dyn}\right\rangle _{E}$
as function of collision energy in Au+Au collisions, solid lines are
for b = 0 fm and dashed lines are for b = 9 fm for corresponding
collisions energy.}
\end{figure}

In this subsection, we give numerical results for the space-averaged
dynamic electric field $\left\langle \text{\textbf{E}}_{dyn}\right\rangle _{E}=\left\langle \sqrt{E_{x-dyn}^{2}+E_{y-dyn}^{2}+E_{z-dyn}^{2}}\right\rangle _{E}$
for Au+Au collisions at intermediate collision energies. We show the
time evolution results for $\left\langle \text{\textbf{E}}_{dyn}\right\rangle _{E}$
in units of $\left(m_{\pi}^{2}\right)$ for collision energies ranging
from 11.5 GeV to 3.5 GeV for two impact parameters i.e, $b=0$ fm
(solid lines) and $b=9$ fm (dashed lines) in Fig. \ref{fig:Space-averaged-dynamical-electri}.
From the figure, we observe that the electric field magnitude is greater
for central collisions $\left(b=0\text{ fm}\right)$ than for non-central
collisions $\left(b=9\text{ fm}\right)$. Similar to the behavior
of the magnetic field, the magnitude of the electric field increases
with collision energy. However, the decay of the electric field is
slower at smaller collision energies, highlighting the enhanced lifetime
of the fields at lower energies. 

As shown in \citep{Taya2024} that the electric fields produced in
central collision at intermediate collisions can provide a unique
and novel opportunity to study Quantum Electrodynamics (QED) in the
non-perturbative regime beyond Schwinger limit i.e., $eE_{cr}:=m_{e}^{2}=\left(0.511\text{ MeV}\right)^{2},$
where $m_{e}$ being the electron mass {[}for details see Ref. \citep{Taya2024}.
Consider the example of the vacuum pair production by strong electric
field, which can be classified as perturbative or non-perturbative
based on two dimenless quantities \citep{Brezin1970,Popov1971,Popov1971a,Dunne2006,Oka2012,Taya2014,Aleksandrov2019}
\begin{equation}
\xi(m)=\frac{eE_{0}T}{m}\text{ and }\nu=eE_{0}T^{2}\label{eq:keydash}
\end{equation}
where $\xi$ is Keldysh parameter \citep{Keldysh1965} or also referred
to as the non-linearity parameter \citep{Fedotov2023}, $m$ is the
mass of the particle to be produced, $E_{0}$ is the field strength
and $T$ is the field lifetime. If $\xi,\nu\gg1$, the pair production
becomes non-perturbative in the sense that the rate of pair production
acquires non-analytical dependencies on $e$ and $E_{0}$ as $\text{exp}\left[-\text{(const)}\times\left(m^{2}/eE_{0}\right)\right]$.
On the other hand, if $\xi,\nu\ll1$ the pair production becomes perturbative
as the rate of pair production has power dependencies in $e$ and
$E_{0}$ as $\propto\left(eE_{0}/m^{2}\right)^{n}$, where $n\in\mathcal{\boldsymbol{N}}$.
From Eq. (\ref{eq:keydash}), it is evident that $E_{0}$ and $T$,
in addition to $m$, determine whether pair production occurs in a
perturbative or non-perturbative regime. While the maximum field strength
$E_{0}$ increases with collision energy, the lifetime $T$ decreases.
To study the non-perturbative regime of QED, we need collision energies
where $E_{0}$ and $T$ are sufficiently large to satisfy $\xi,\nu\gg1$. 

\begin{figure*}
\includegraphics[width=8cm]{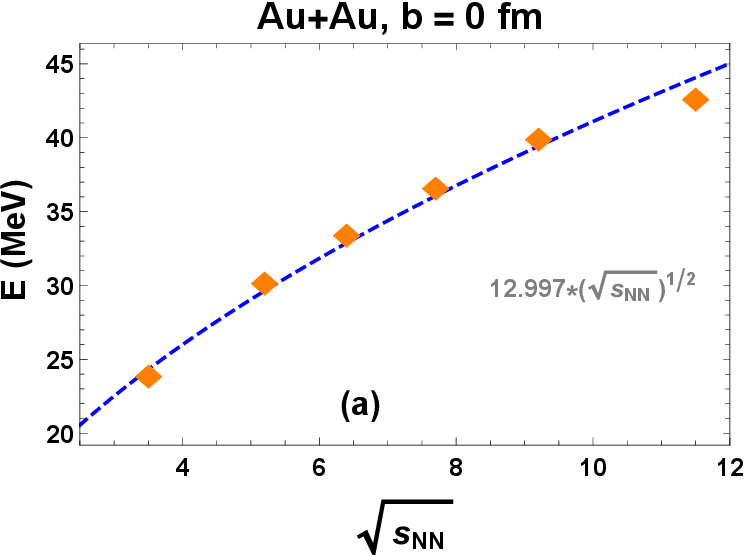}\includegraphics[width=8cm]{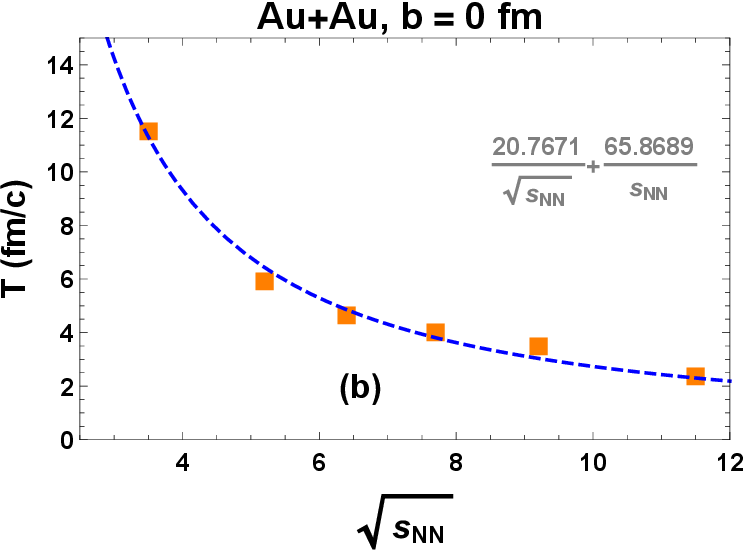}

\caption{\label{fig:Peak-strength-of}Peak strength of $\left\langle \text{\textbf{E}}_{dyn}\right\rangle _{E}$
as function of collision energy for Au+Au collisions at $b=0$ fm
in left panel. Right panel shows the effective time for $\left\langle \text{\textbf{E}}_{dyn}\right\rangle _{E}$
as a function of collision energy for Au+Au collision at $b=0$ fm.}
\end{figure*}

In Fig. (\ref{fig:Peak-strength-of}a), we show the peak magnitude
of electric field $_{max}\left\langle \text{\textbf{E}}_{dyn}\right\rangle _{E}$
(MeV scale ) as a function of collision energy. The maximum peak strength
ranges from 20 MeV to 45 MeV for collision energies between 3.5 GeV
to 11.5 GeV. Although the peak field strengths are weaker compared
to those at higher collision energies, they are still significantly
stronger than the Schwinger limit. The relationship between the peak
strength and collision energy is well-fitted by the formula:

\begin{equation}
_{max}\left\langle \text{\textbf{E}}_{dyn}\right\rangle _{E}=12.997\times\left(\frac{\sqrt{s_{NN}}}{1\text{GeV}}\right)^{1/2}.\label{eq:fitE}
\end{equation}
In Fig. (\ref{fig:Peak-strength-of}b), we give the effective lifetime
the electric field, calculated as \citep{Taya2024} 
\begin{equation}
T=\int_{E(t)>_{max}\left\langle \text{\textbf{E}}_{dyn}\right\rangle _{E}/2}dt.
\end{equation}
The results show that $T$ increases as the collision energy decreases
and the best fitting curve for $T$ is 
\begin{equation}
T=20.7671\times\left(\frac{\sqrt{s_{NN}}}{1\text{ GeV}}\right)^{-1}+65.8689\times\left(\frac{\sqrt{s_{NN}}}{1\text{ GeV}}\right)^{-2}.\label{eq:fitT}
\end{equation}
Our fitting results shown in Eqs. (\ref{eq:fitE}) and (\ref{eq:fitT})
have different coefficients from the fitting results shown in \citep{Taya2024}
due to the differences in the transport model and simulation methodology.
However, the qualitative behavior remains consistent i.e., at lower
collision energies, the electric fields are significantly stronger
relative to the Schwinger limit and exhibit much longer lifetimes.
This delayed lifetime behavior is attributed to contributions from
internal fields (influenced by conductivities) and baryon stopping
at lower collision energies. Similar to \citep{Taya2024}, our results
also indicate that heavy-ion collisions at lower collision energies
can provide an excellent opportunity to study the non-perturbative
regime of QED. The combination of strong electric fields (beyond the
Schwinger limit) and extended lifetimes may allows for the exploration
of non-linearity parameters $\xi,\nu\gg1$ that lead to non-perturbative
dependencies. This opens a new avenue for studying fundamental QED
processes in extreme conditions.

\subsubsection{Dynamic electromagnetic anomaly}

\begin{figure*}
\fbox{\begin{minipage}[t]{7.6cm}%
\includegraphics[width=7.5cm]{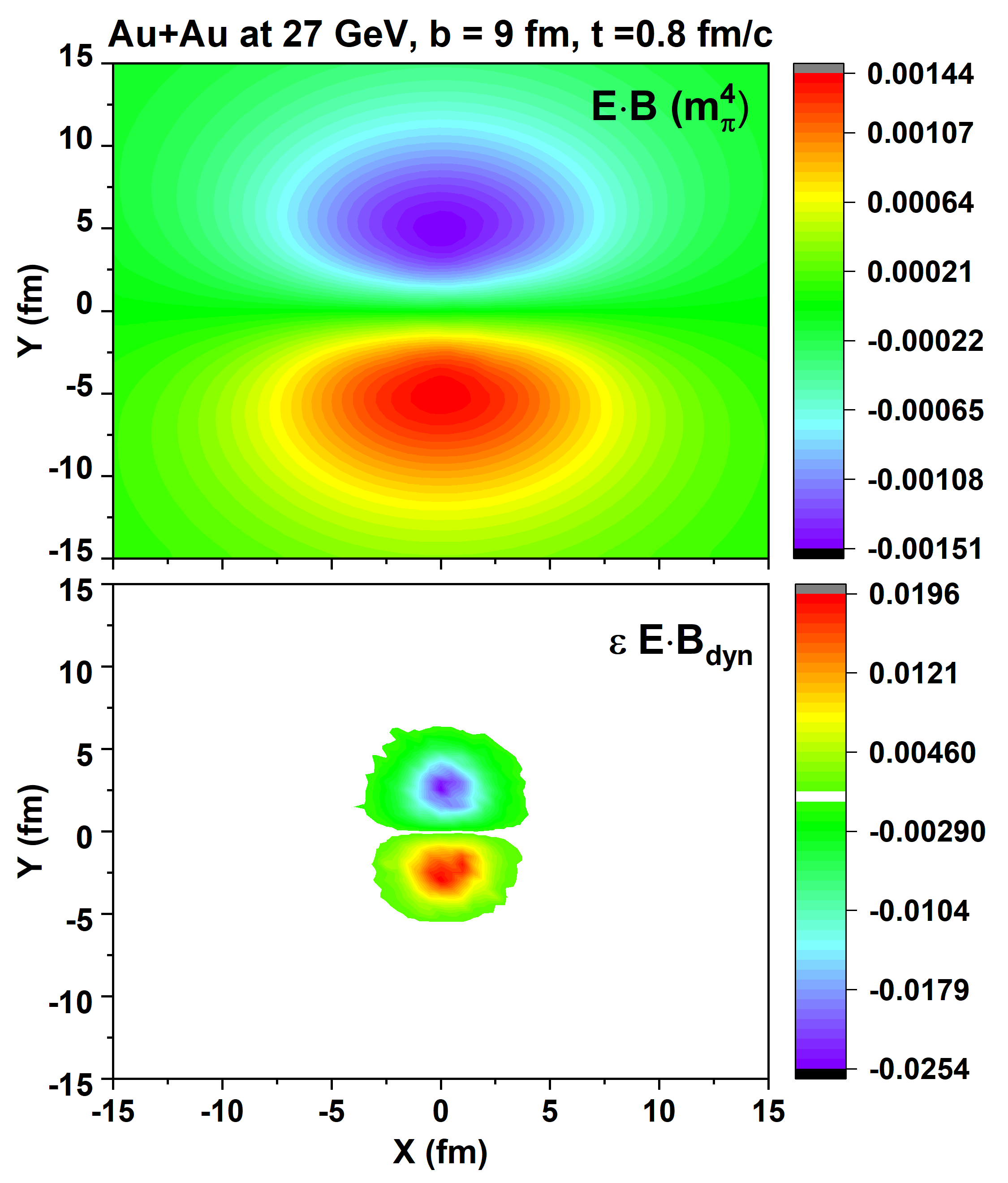}

A%
\end{minipage}}%
\fbox{\begin{minipage}[t]{7.6cm}%
\includegraphics[width=7.5cm]{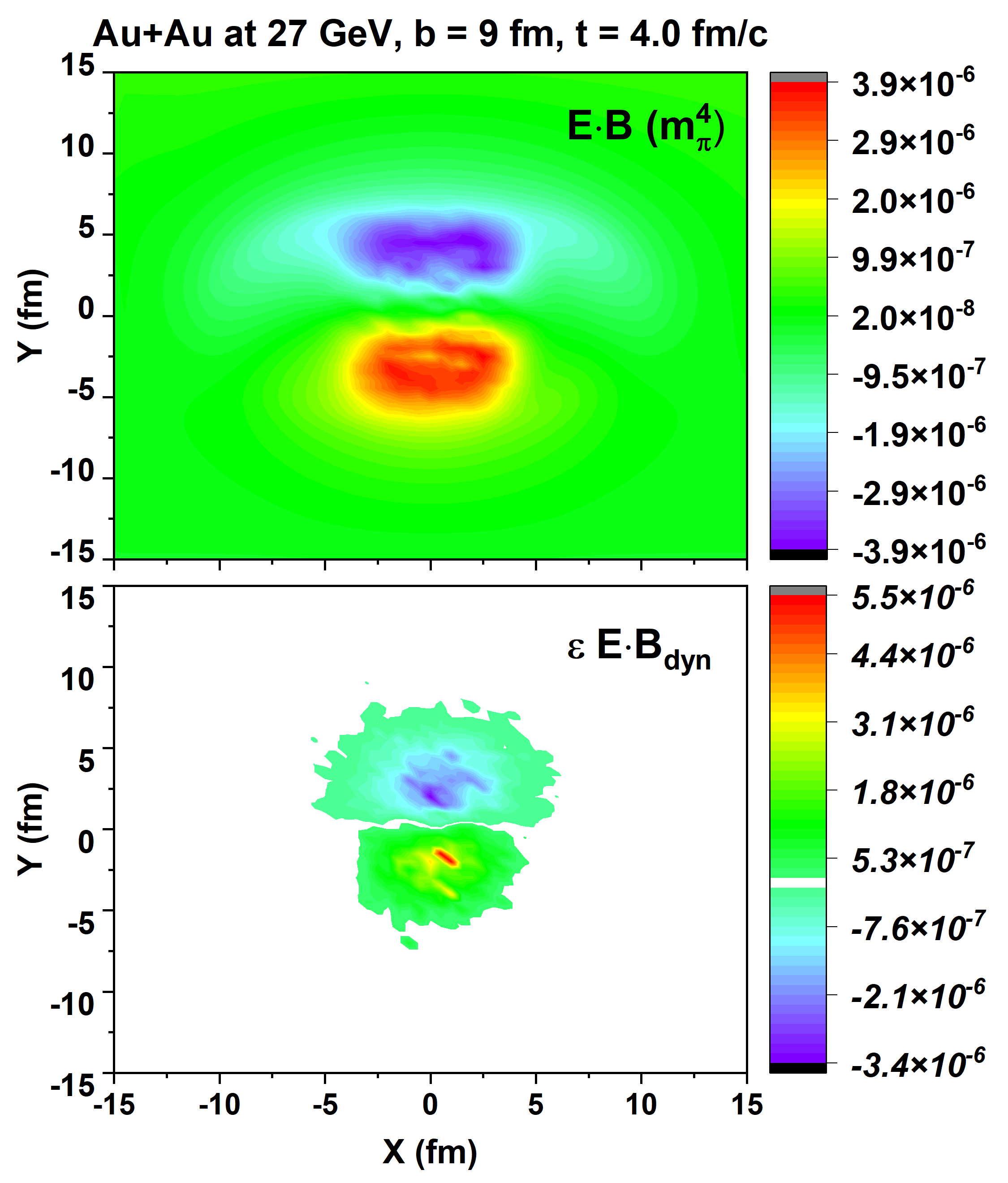}

B%
\end{minipage}}\caption{\label{fig:contou-EBxy}Spatial distribution of dynamical electromagnetic anomaly
in transverse plane in Au+Au collisions at 27 GeV and $b=9$ fm at $t=0.8$ fm/c
(panel A) and $t=4$ fm/c (panel B). Lower plots of both panel shows spatial distribution for the product of energy density and dynamical electromagnetic anomaly.}
\end{figure*}

In this section, we investigate the dynamical electromagnetic anomaly
$\boldsymbol{\textbf{E}}\cdot\boldsymbol{\textbf{B}}$ in heavy ion collisions.
The dynamical electromagnetic anomaly is calculated as 
\begin{equation}
\left(\text{\textbf{E}}\cdot\boldsymbol{\textbf{B}}\right)_{dyn}=E_{x,tot}B_{x,tot}+E_{y,tot}B_{y,tot}+E_{z,tot}B_{z,tot},
\end{equation}
where $F_{i,tot}=F_{i,ext}+F_{i,int}$, with $F$ being $E$ or $B$.
The spatial distribution of $\text{\textbf{E}}\cdot\boldsymbol{\textbf{B}}$
for Au+Au collisions at $\sqrt{s_{NN}}=27$ GeV and $b=9$ fm in transverse plane
is given in Fig. \ref{fig:contou-EBxy}. Similar to strategy adopted
in subsection \ref{subsec:Spatial-Distributions:}, there are two
snapshots at different time are presented in panel A and panel B belonging
to earlier time $\left(t=0.8\text{ fm/c}\right)$ and later time $\left(t=4\text{ fm/c}\right)$
respectively of evolving system. In each panel upper plot shows the spatial
distribution for $\text{\textbf{E}}\cdot\boldsymbol{\textbf{B}}$ and lower plot shows
the results is for $\varepsilon\left(\text{\textbf{E}}\cdot\boldsymbol{\textbf{B}}\right)$,
where $\varepsilon$ is energy density. From the spatial distribution
in the upper plots of both panels A and B of Fig. \ref{fig:contou-EBxy},
we observe that the electromagnetic anomaly exhibits a dipolar structure.
This dipolar structure is symmetric when flipping the sign of the
$x$-coordinate but asymmetric when flipping the sign of the $y$-coordinate.
In lower plot of each panel of Fig. \ref{fig:contou-EBxy}, spatial
distribution of $\text{\textbf{E}}\cdot\boldsymbol{\textbf{B}}$ times
energy density $\left(\varepsilon\right)$ is shown, which also shows
a dipolar structure. From the spatial distribution one can notice
that directly calculating the dynamical electromagnetic anomaly weighted
by energy density will result in $\left\langle \text{\textbf{E}}\cdot\boldsymbol{\textbf{B}}\right\rangle _{E}\approx0$,
however, when averaging in upper $\left(y>0\right)$ and/or lower
$\left(y<0\right)$ halves of spatial plane, the results are non-zero.

\begin{figure*}
\includegraphics[width=8cm]{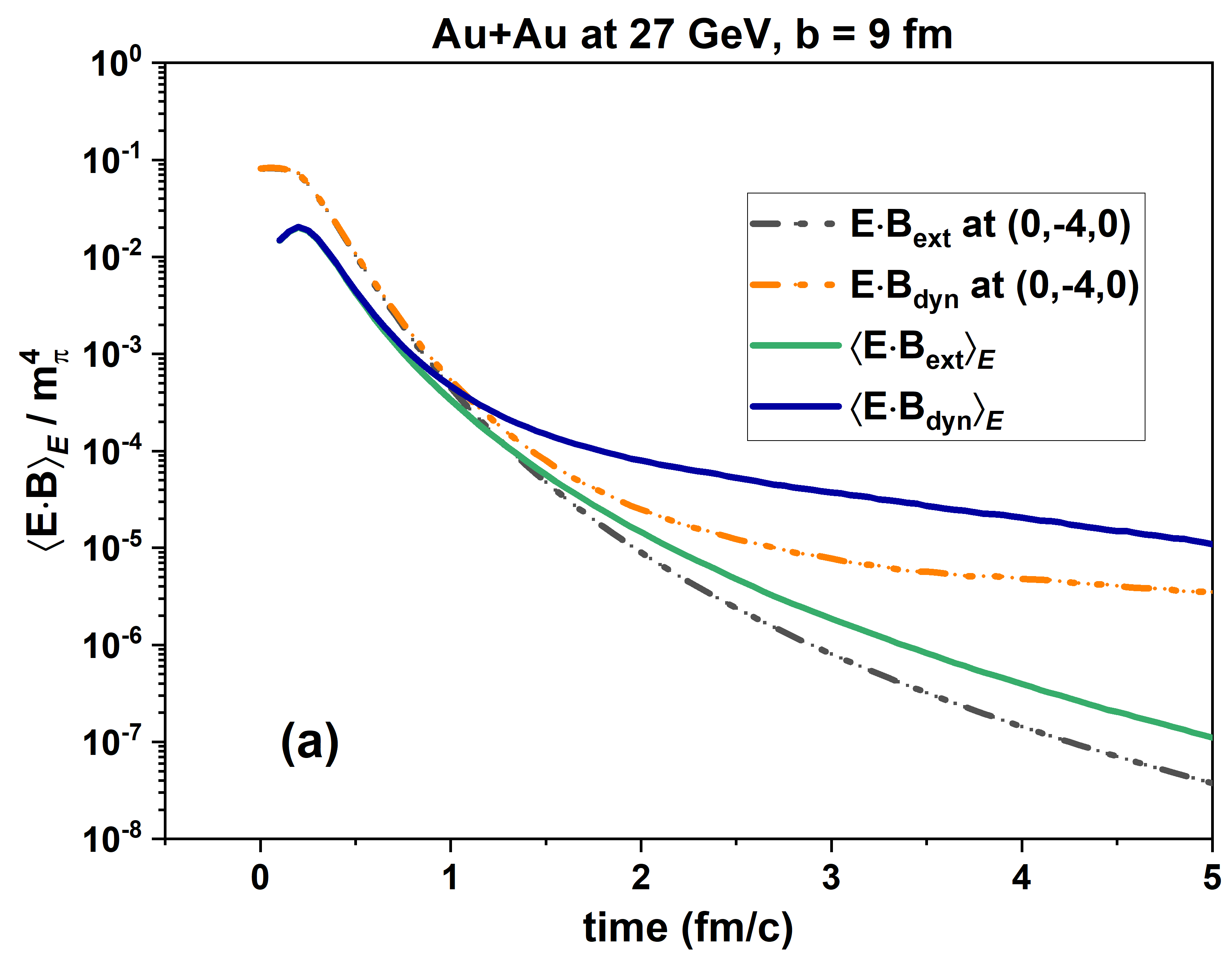}\includegraphics[width=8cm]{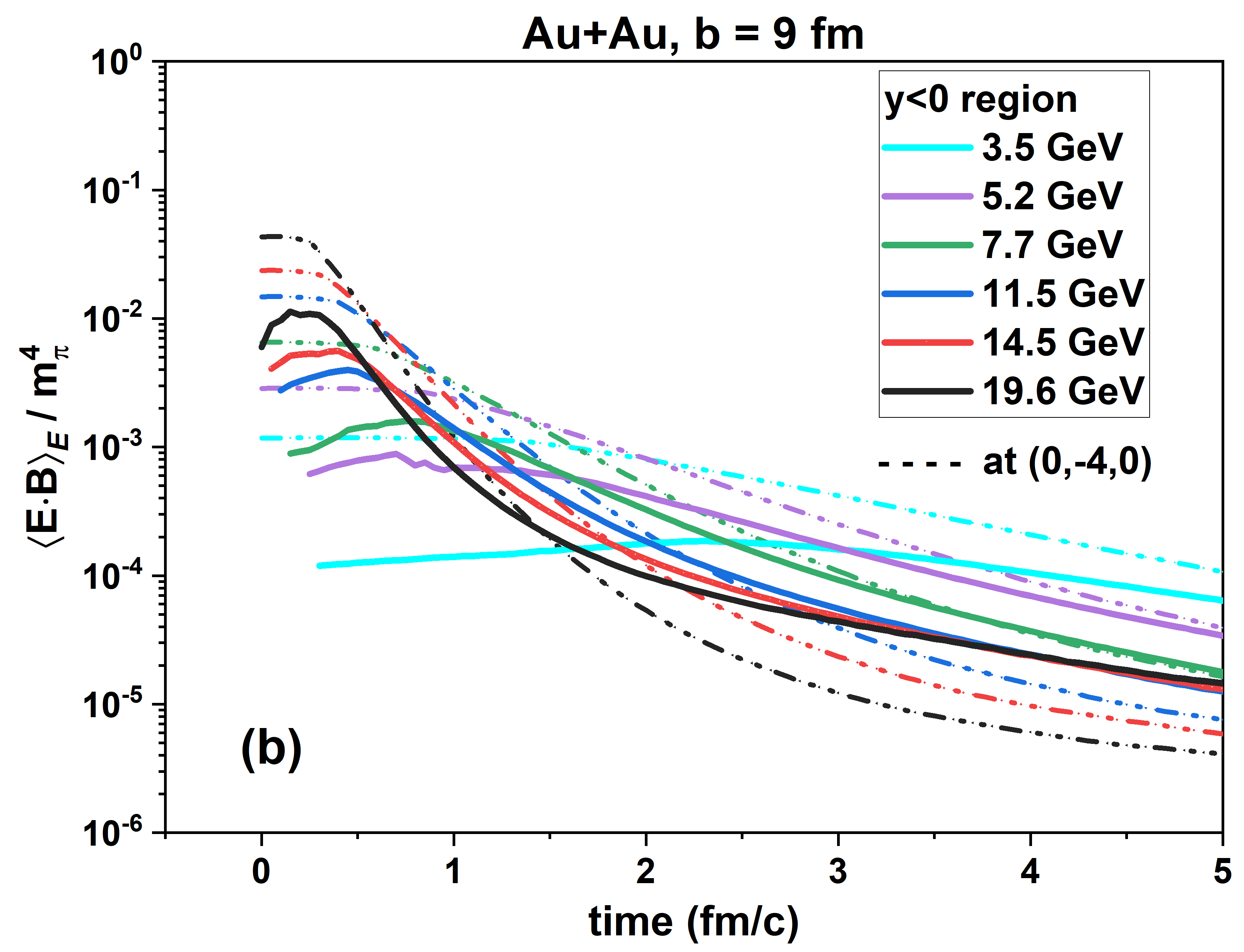}

\caption{\label{fig:Time-Evolution-ofEB}
Time evolution of the dynamical electromagnetic anomaly in Au+Au collisions at $\sqrt{s_{NN}}=27$ GeV and $b=9$ fm is shown in left panel. Right panel shows the time evolution of electromagnetic anomaly for other intermediate collision energies.}
\end{figure*}

In Fig. (\ref{fig:Time-Evolution-ofEB}a), we show time evolution
of $\left\langle \text{\textbf{E}}\cdot\boldsymbol{\textbf{B}}\right\rangle _{E}$
in lower half space i.e., $y<0$ region in Au+Au collisions at $\sqrt{s_{NN}}=27$
GeV using formula given in Eq. (\ref{eq:space-averaged}). We give a
comparison between $\left\langle \left(\text{\textbf{E}}\cdot\boldsymbol{\textbf{B}}\right)_{ext}\right\rangle _{E}$,
$\left\langle \left(\text{\textbf{E}}\cdot\boldsymbol{\textbf{B}}\right)_{dyn}\right\rangle _{E}$
,$\left(\text{\textbf{E}}\cdot\boldsymbol{\textbf{B}}\right)_{ext}$
at $(0,-4,0)$ and $\left(\text{\textbf{E}}\cdot\boldsymbol{\textbf{B}}\right)_{dyn}$
at $(0,-4,0)$ fm. A visible difference can be seen between $\left\langle \left(\text{\textbf{E}}\cdot\boldsymbol{\textbf{B}}\right)_{dyn}\right\rangle _{E}$
(solid blue line) and $\left\langle \left(\text{\textbf{E}}\cdot\boldsymbol{\textbf{B}}\right)_{ext}\right\rangle _{E}$
(solid green line), we see that initially the magnitude for dynamical
fields i.e., $\left\langle \left(\text{\textbf{E}}\cdot\boldsymbol{\textbf{B}}\right)_{dyn}\right\rangle _{E}$
and $\left(\text{\textbf{E}}\cdot\boldsymbol{\textbf{B}}\right)_{dyn}$ at
$(0,-4,0)$ has smaller magnitude but damps slower than the $\left\langle \left(\text{\textbf{E}}\cdot\boldsymbol{\textbf{B}}\right)_{ext}\right\rangle _{E}$
and $\left(\text{\textbf{E}}\cdot\boldsymbol{\textbf{B}}\right)_{ext}$ 
at $(0,-4,0)$. Notably, the decay of $\left\langle \left(\text{\textbf{E}}\cdot\boldsymbol{\textbf{B}}\right)_{dyn}\right\rangle _{E}$
is the slowest among all. The time evolution of $\left\langle \left(\text{\textbf{E}}\cdot\boldsymbol{\textbf{B}}\right)_{dyn}\right\rangle _{E}$
(solid lines) and $\left(\text{\textbf{E}}\cdot\boldsymbol{\textbf{B}}\right)_{dyn}$ at
$(0,-4,0)$ for other collision energies is shown in Fig (\ref{fig:Time-Evolution-ofEB}b),
where we see that, as collision energy decreases, the peak magnitude
of $\left\langle \left(\text{\textbf{E}}\cdot\boldsymbol{\textbf{B}}\right)_{dyn}\right\rangle _{E}$
decreases, but the decay rate becomes more slower at lower collision
energies. 

Given the spatial and temporal variation of electromagnetic anomaly
in previous figures, their overall effects on physical observable
should be evaluated at an average level over the full volume
and lifetime of quark and nuclear matter. To quantify net effect by
dynamical electromagnetic anomaly we define space-time averaged dynamical
electromagnetic anomaly as 
\begin{equation}
\left\langle \left(\text{\textbf{E}}\cdot\boldsymbol{\textbf{B}}\right)_{dyn}\right\rangle _{TE}=\frac{\int dt\left\langle \left(\text{\textbf{E}}\cdot\boldsymbol{\textbf{B}}\right)_{dyn}\right\rangle _{E}}{\int dt},
\end{equation}
 since we have discretized the whole time period so above equation
can be transformed into 
\begin{equation}
\left\langle \left(\text{\textbf{E}}\cdot\boldsymbol{\textbf{B}}\right)_{dyn}\right\rangle _{TE}\equiv\frac{\sum_{i}\left\langle \left(\text{\textbf{E}}\cdot\boldsymbol{\textbf{B}}\right)_{i,dyn}\right\rangle _{E}\Delta t_{i}}{\sum_{i}\Delta t_{i}}
\end{equation}
where $\left\langle \left(\text{\textbf{E}}\cdot\boldsymbol{\textbf{B}}\right)_{i,dyn}\right\rangle _{E}$
is space-averaged dynamical electromagnetic anomaly at $i$-th time
step. As it is shown in Fig (\ref{fig:Time-Evolution-ofEB}a) that
initially magnitude of $\left\langle \left(\text{\textbf{E}}\cdot\boldsymbol{\textbf{B}}\right)_{dyn}\right\rangle _{E}$
decays faster but this decay becomes slower once the contribution
from induced part of electric and magnetic field components starts
playing role and after $t=2$ fm/c the decay become rate becomes very
slow. Moreover magnitude at $t=2$ fm/c is much smaller than initial
time so we take time average interval $0\rightarrow2$ fm/c while
calculating $\left\langle \left(\text{\textbf{E}}\cdot\boldsymbol{\textbf{B}}\right)_{dyn}\right\rangle _{TE}$. 

\begin{figure}
\includegraphics[width=8cm]{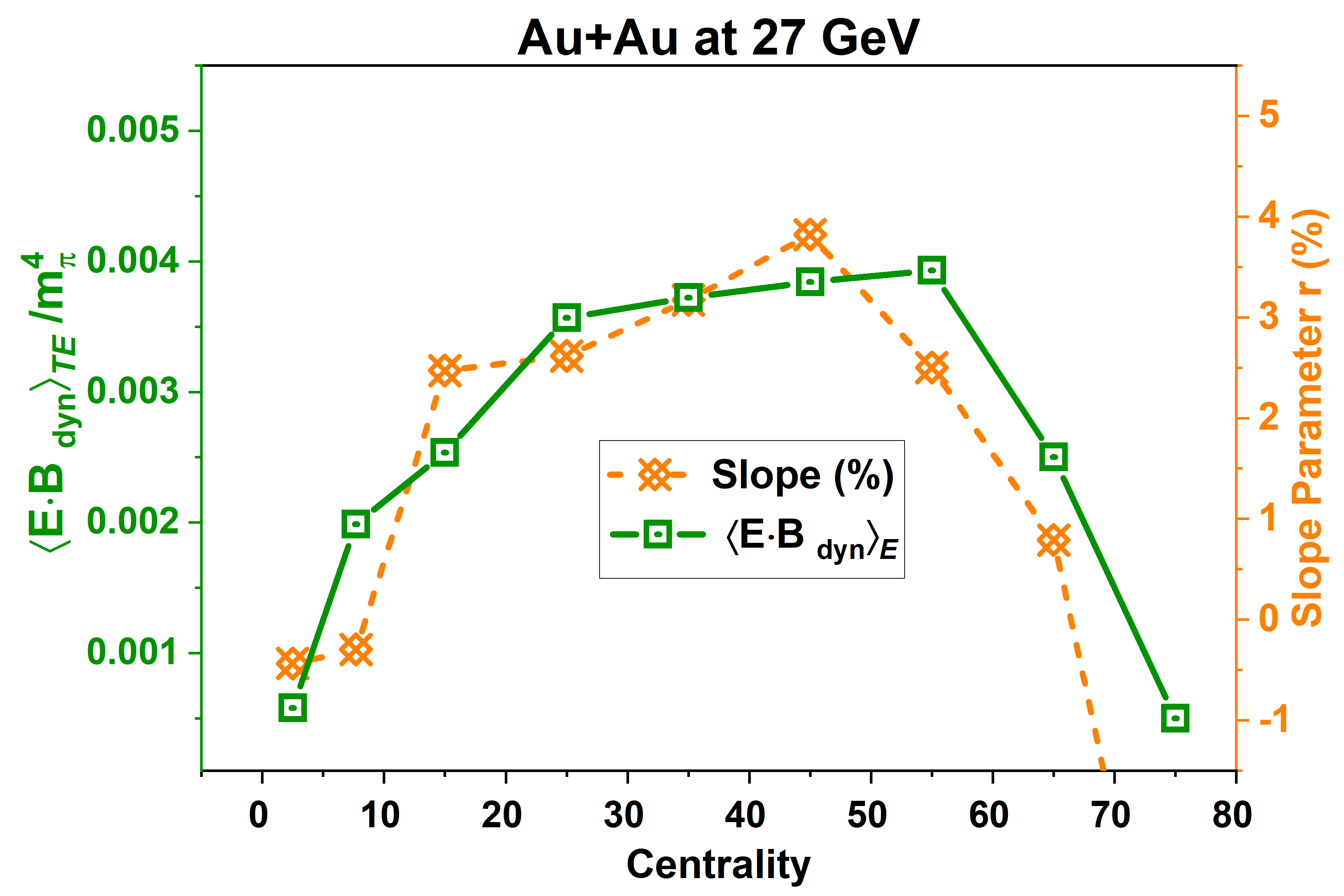}

\caption{\label{fig:Slope-vs-EB}The space-time averaged dynamical electromagnetic
anomaly i.e, $\left\langle \left(\boldsymbol{E}\cdot\boldsymbol{B}\right)_{dyn}\right\rangle _{TE}$
and the slope parameter $r$ as functions of centrality in Au + Au
collisions at 27 GeV. }

\end{figure}

In Fig. \ref{fig:Slope-vs-EB}, we present the space-time averaged
magnitude of dynamical electromagnetic anomaly i.e., $\left\langle \left(\text{\textbf{E}}\cdot\boldsymbol{\textbf{B}}\right)_{dyn}\right\rangle _{TE}$
for different centralities at 27 GeV and compare them to the slope
parameter $r$ derived from charge-dependent elliptic flow differences
for charged pions, as reported by the STAR collaboration in \citep{STAR:2015wza}.
The slope parameter $r$, defined as $v_{2}\left(\pi^{\pm}\right)=v_{2}^{\text{base}}\left(\pi^{\pm}\right)\mp rA_{ch}/2$,
with $A_{ch}=\left(N_{+}-N_{-}\right)/\left(N_{+}+N_{-}\right)$
characterizing the charge asymmetry, quantifies the difference in
elliptic flow $v_{2}$ between $\pi^{+}$ and $\pi^{-}$. The similarity
in trends between the space-time averaged dynamical electromagnetic
anomaly $\left\langle \left(\text{\textbf{E}}\cdot\boldsymbol{\textbf{B}}\right)_{dyn}\right\rangle _{TE}$
and the measured $r$ parameter as a function of centrality, as shown
in the Fig. \ref{fig:Slope-vs-EB}, suggests that the QED anomaly
$\text{\textbf{E}}\cdot\boldsymbol{\textbf{B}}$ could also be a potential
mechanism for the $v_{2}$ separation observed between positive and
negative charges in Au+Au collisions at 27 GeV. 

\section{Summary and outlook}

\label{sec:summary}

In this study, we have provided a comprehensive analysis of the dynamical
electromagnetic fields and dynamical electromagnetic anomaly in heavy-ion
collisions, focusing on intermediate collision energies. Using the
UrQMD transport model, which accounts for full collision process,
we calculated the current density and numerically solve Maxwell's
equations with finite conductivities to obtain dynamical fields. The
research conducted in this papers explores the space-time evolution
of dynamical electromagnetic fields and electromagnetic anomaly in
the presence of a medium i.e., quark-gluon plasma (QGP). Moreover,
space-averaged fields weighted by energy density were introduced to
mitigate over- or underestimations in local field-induced effects. 

While studying the dynamical magnetic fields, at intermediate collision
energies, we show that they exhibit extended lifetimes and slower
decay due to QGP contributions$\left(\sigma\text{ and }\sigma_{\chi}\right)$,
which may enable a prolonged influence on field-induced phenomena.
The peak magnetic field strength increases with collision energy,
but at lower collision energies, the fields decay more slowly and
persist longer, maintaining a larger magnitude at later times. We
also study the impact of the magnetic field on the spin polarization
of $\Lambda$ and $\bar{\Lambda}$. We found that the field strength
at later times is insufficient to induce a significant splitting in
their global polarization. This result is consistent with previous
studies and experimental data from the STAR collaboration. 

Dynamical electric fields exhibit behavior similar to magnetic fields,
i.e., they exhibit longer lifetimes and slower decay when medium feedback
is considered in terms of conductivities. We also show that the peak
electric field strength increases with collision energy, but the lifetime
is significantly extended at lower collision energies, which poses
the potential of dynamical electric fields in central collisions to
probe the non-perturbative regime of QED beyond the Schwinger limit.
For collision energies between 3.5 GeV and 11.5 GeV, the maximum electric
field strength was found to range from 20 MeV to 45 MeV. While smaller
than fields at higher collision energies, these values remain critical
for studying the Schwinger effect due to their magnitude and extended
lifetime. Our results suggest that intermediate-energy heavy-ion collisions
generate non-perturbatively strong electric fields capable of inducing
the Schwinger effect. Our results are qualitatively consistent with
previous studies. The collisions at intermediate energies may provide
a experimental setup to test this phenomenon. However, future work
needs to focus and predict the potential experimental signatures of
the Schwinger effect in heavy-ion collisions.

Furthermore, we extended our work to study dynamical electromagnetic
anomaly. We analyzed the spatial distribution and time evolution
of dynamical electromagnetic anomaly. We define and calculate space-time
averaged dynamical electromagnetic anomaly and compare them to the
slope parameter $r$. These dependencies are found qualitatively consistent
with the STAR data on the slope parameter $r$ as a function of the
centrality, indicating the QED anomaly could be an potential mechanism
that drives the $v_{2}$ separation between positive and negative charges.

In the calculations of this paper, we have used finite conductivities,
one can improve by introducing time dependent conductivities, which
is expected to further effect the electromagnetic fields and deserves
a detailed study in the future. 
Our study, consistent with previous researches, suggests that intermediate-energy heavy-ion collisions provide a unique opportunity to explore QCD under extreme conditions characterized by strong electric fields. While observing significant effects at the hadronic scale may be challenging, nonperturbative changes are more likely to occur in the deconfined phase of QCD. Numerous studies have investigated QCD in the presence of strong magnetic fields, predicting significant modifications to the QCD phase diagram . However, studies addressing strong electric fields are comparatively sparse and no consensus has been reached regarding their effects. This highlights the need for further theoretical investigations to understand the impact of strong electric fields on QCD and its phase structure.

\begin{acknowledgments}
We thank X.G. Huang, G.Torrieri and Q. Wang for helpful
discussions.  I.Siddique is supported by the Ministry of Science and Technology (MOST) of China under Grant No. QN2023205001L and RFIS-NSFC under Grant No. 12350410364. M. Huang and A. Huang are supported by the National Natural Science Foundation of China (NSFC) under Grant No. 12235016 and 12221005. A. Huang is also grateful for the support from the NSFC under Grant No. 12205309.
\end{acknowledgments}

\bibliographystyle{h-physrev5}
\bibliography{SCrefs}

\begin{thebibliography}{10}

\bibitem{Kharzeev:2007jp}
D.~E. Kharzeev, L.~D. McLerran, and H.~J. Warringa,
\newblock Nucl. Phys. A {\bf 803}, 227 (2008), arXiv:0711.0950.

\bibitem{Skokov2009}
V.~Skokov, A.~Y. Illarionov, and V.~Toneev,
\newblock Int. J. Mod. Phys. A {\bf 24}, 5925 (2009), arXiv:0907.1396.

\bibitem{Asakawa2010}
M.~Asakawa, A.~Majumder, and B.~Muller,
\newblock Physical Review C {\bf 81}, 064912 (2010).

\bibitem{Bzdak:2011yy}
A.~Bzdak and V.~Skokov,
\newblock Phys. Lett. B {\bf 710}, 171 (2012), arXiv:1111.1949.

\bibitem{Deng:2012pc}
W.-T. Deng and X.-G. Huang,
\newblock Phys. Rev. C {\bf 85}, 044907 (2012), arXiv:1201.5108.

\bibitem{Bloczynski:2012en}
J.~Bloczynski, X.-G. Huang, X.~Zhang, and J.~Liao,
\newblock Phys. Lett. B {\bf 718}, 1529 (2013), arXiv:1209.6594.

\bibitem{Tuchin:2013apa}
K.~Tuchin,
\newblock Phys. Rev. C {\bf 88}, 024911 (2013), arXiv:1305.5806.

\bibitem{Tuchin:2014iua}
K.~Tuchin,
\newblock Phys. Rev. C {\bf 91}, 064902 (2015), arXiv:1411.1363.

\bibitem{Roy2015}
V.~Roy and S.~Pu,
\newblock Physical Review C {\bf 92}, 064902 (2015).

\bibitem{STAR:2021mii}
STAR, M.~Abdallah {\em et~al.},
\newblock Phys. Rev. C {\bf 105}, 014901 (2022), arXiv:2109.00131.

\bibitem{STARCollaboration2022}
{STAR Collaboration} {\em et~al.},
\newblock Physics Letters B {\bf 839}, 137779 (2022), arXiv:2209.03467.

\bibitem{STARCollaboration2017}
{STAR Collaboration},
\newblock Nature 548, 62 (2017) {\bf 548}, 62 (2017), arXiv:1701.06657.

\bibitem{Kharzeev:2015znc}
D.~E. Kharzeev, J.~Liao, S.~A. Voloshin, and G.~Wang,
\newblock Prog. Part. Nucl. Phys. {\bf 88}, 1 (2016), arXiv:1511.04050.

\bibitem{Li2020}
W.~Li and G.~Wang,
\newblock Annual Review of Nuclear and Particle Science {\bf 70}, 293 (2020).

\bibitem{Huang2016}
X.-G. Huang,
\newblock Reports on Progress in Physics {\bf 79}, 076302 (2016).

\bibitem{Zhang:2022lje}
J.-J. Zhang {\em et~al.},
\newblock Phys. Rev. Res. {\bf 4}, 033138 (2022), arXiv:2201.06171.

\bibitem{Sheng:2017lfu}
X.-l. Sheng, D.~H. Rischke, D.~Vasak, and Q.~Wang,
\newblock Eur. Phys. J. A {\bf 54}, 21 (2018), arXiv:1707.01388.

\bibitem{Sheng:2018jwf}
X.-L. Sheng, R.-H. Fang, Q.~Wang, and D.~H. Rischke,
\newblock Phys. Rev. D {\bf 99}, 056004 (2019), arXiv:1812.01146.

\bibitem{Fukushima:2015wck}
K.~Fukushima, K.~Hattori, H.-U. Yee, and Y.~Yin,
\newblock Phys. Rev. D {\bf 93}, 074028 (2016), arXiv:1512.03689.

\bibitem{Hattori:2017qih}
K.~Hattori, X.-G. Huang, D.~H. Rischke, and D.~Satow,
\newblock Phys. Rev. D {\bf 96}, 094009 (2017), arXiv:1708.00515.

\bibitem{Lin:2021sjw}
S.~Lin and L.~Yang,
\newblock JHEP {\bf 06}, 054 (2021), arXiv:2103.11577.

\bibitem{Peng:2023rjj}
H.-H. Peng, X.-L. Sheng, S.~Pu, and Q.~Wang,
\newblock Phys. Rev. D {\bf 107}, 116006 (2023), arXiv:2304.00519.

\bibitem{Siddique2022}
I.~Siddique, S.~Cao, U.~Tabassam, M.~Saeed, and M.~Waqas,
\newblock Phys. Rev. C {\bf 105}, 054909 (2022), arXiv:2201.09634.

\bibitem{Bass:1998ca}
S.~A. Bass {\em et~al.},
\newblock Prog. Part. Nucl. Phys. {\bf 41}, 225 (1998), nucl-th/9803035.

\bibitem{Bleicher:1999xi}
M.~Bleicher {\em et~al.},
\newblock J. Phys. {\bf G25}, 1859 (1999), hep-ph/9909407.

\bibitem{Lin:2004en}
Z.-W. Lin, C.~M. Ko, B.-A. Li, B.~Zhang, and S.~Pal,
\newblock Phys. Rev. {\bf C72}, 064901 (2005), arXiv:nucl-th/0411110.

\bibitem{Lin2021}
Z.-W. Lin and L.~Zheng,
\newblock Nuclear Science and Techniques {\bf 32} (2021).

\bibitem{Hattori2017}
K.~Hattori and X.-G. Huang,
\newblock Nuclear Science and Techniques {\bf 28} (2017).

\bibitem{McLerran:2013hla}
L.~McLerran and V.~Skokov,
\newblock Nucl. Phys. A {\bf 929}, 184 (2014), arXiv:1305.0774.

\bibitem{Li:2016tel}
H.~Li, X.-l. Sheng, and Q.~Wang,
\newblock Phys. Rev. C {\bf 94}, 044903 (2016), arXiv:1602.02223.

\bibitem{Siddique:2019gqh}
I.~Siddique, R.-j. Wang, S.~Pu, and Q.~Wang,
\newblock Phys. Rev. D {\bf 99}, 114029 (2019), arXiv:1904.01807.

\bibitem{Siddique:2021smf}
I.~Siddique, X.-L. Sheng, and Q.~Wang,
\newblock Phys. Rev. C {\bf 104}, 034907 (2021), arXiv:2106.00478.

\bibitem{Anping2023}
A.~Huang, D.~She, S.~Shi, M.~Huang, and J.~Liao,
\newblock Physical Review C {\bf 107}, 034901 (2023).

\bibitem{Fermi1950}
E.~Fermi,
\newblock Progress of Theoretical Physics {\bf 5}, 570 (1950).

\bibitem{Landau:1953gs}
L.~D. Landau,
\newblock Izv. Akad. Nauk Ser. Fiz. {\bf 17}, 51 (1953).

\bibitem{Bjorken:1982qr}
J.~D. Bjorken,
\newblock Phys. Rev. {\bf D27}, 140 (1983).

\bibitem{Yee1966}
K.~Yee,
\newblock IEEE Transactions on Antennas and Propagation {\bf 14}, 302 (1966).

\bibitem{Zhao2019a}
X.-L. Zhao, G.-L. Ma, and Y.-G. Ma,
\newblock Phys. Rev. C {\bf 99}, 034903 (2019), arXiv:1901.04151.

\bibitem{Zhao:2019ybo}
X.-L. Zhao, G.-L. Ma, and Y.-G. Ma,
\newblock Phys. Lett. B {\bf 792}, 413 (2019), arXiv:1901.04156.

\bibitem{Ding2011}
H.-T. Ding {\em et~al.},
\newblock Physical Review D {\bf 83}, 034504 (2011).

\bibitem{Aarts2015}
G.~Aarts {\em et~al.},
\newblock Journal of High Energy Physics {\bf 2015} (2015).

\bibitem{Siddique2024}
I.~Siddique and U.~Tabassam,
\newblock Physical Review C {\bf 109}, 034905 (2024).

\bibitem{Tuchin2013}
K.~Tuchin,
\newblock Advances in High Energy Physics {\bf 2013}, 1 (2013).

\bibitem{Li2023}
H.~Li, X.-L. Xia, X.-G. Huang, and H.~Z. Huang,
\newblock Physical Review C {\bf 108}, 044902 (2023).

\bibitem{Becattini2017}
F.~Becattini, I.~Karpenko, M.~A. Lisa, I.~Upsal, and S.~A. Voloshin,
\newblock Physical Review C {\bf 95}, 054902 (2017).

\bibitem{Mueller2018}
B.~Muller and A.~Schafer,
\newblock Physical Review D {\bf 98}, 071902 (2018).

\bibitem{Peng2023}
H.-H. Peng, S.~Wu, R.-j. Wang, D.~She, and S.~Pu,
\newblock Physical Review D {\bf 107}, 096010 (2023).

\bibitem{STAR2023}
M.~I. Abdulhamid {\em et~al.},
\newblock Physical Review C {\bf 108}, 014910 (2023).

\bibitem{Taya2024}
H.~Taya, T.~Nishimura, and A.~Ohnishi,
\newblock Physical Review C {\bf 110}, 014901 (2024).

\bibitem{Brezin1970}
E.~Brezin and C.~Itzykson,
\newblock Physical Review D {\bf 2}, 1191 (1970).

\bibitem{Popov1971}
V.~S. Popov,
\newblock Zh. Eksp. Teor. Fiz. {\bf 61}, 1334 (1971).

\bibitem{Popov1971a}
V.~S. Popov,
\newblock JETP Lett. {\bf 13}, 185 (1971).

\bibitem{Dunne2006}
G.~V. Dunne, Q.-h. Wang, H.~Gies, and C.~Schubert,
\newblock Physical Review D {\bf 73}, 065028 (2006).

\bibitem{Oka2012}
T.~Oka,
\newblock Physical Review B {\bf 86}, 075148 (2012).

\bibitem{Taya2014}
H.~Taya, H.~Fujii, and K.~Itakura,
\newblock Physical Review D {\bf 90}, 014039 (2014).

\bibitem{Aleksandrov2019}
I.~A. Aleksandrov, G.~Plunien, and V.~M. Shabaev,
\newblock Physical Review D {\bf 99}, 016020 (2019).

\bibitem{Keldysh1965}
L.~V. Keldysh,
\newblock J. Exp. Theor. Phys. {\bf 20}, 1307 (1965).

\bibitem{Fedotov2023}
A.~Fedotov {\em et~al.},
\newblock Physics Reports {\bf 1010}, 1 (2023).

\bibitem{STAR:2015wza}
STAR, L.~Adamczyk {\em et~al.},
\newblock Phys. Rev. Lett. {\bf 114}, 252302 (2015), arXiv:1504.02175.

\end{thebibliography}

\end{document}